\documentclass[aps,prb,twocolumn,superscriptaddress]{revtex4-2}
\usepackage{graphicx}
\usepackage{mathrsfs}
\usepackage[colorlinks=true,allcolors=blue]{hyperref}
\usepackage{bm}
\usepackage{amsmath}
\usepackage{dcolumn}
\usepackage{epstopdf}
\usepackage{dsfont}
\usepackage{amssymb}
\usepackage{tabularx}
\usepackage{array}
\usepackage{colordvi}
\usepackage{braket}
\usepackage{float}
\usepackage{mathtools}
\usepackage{upgreek}
\usepackage{booktabs}
\usepackage{multirow}

\begin{document}
\title{Voltage-tunable giant nonvolatile multiple-state resistance in interlayer-sliding ferroelectric \emph{h}-BN engineered van der Waals multiferroic tunnel junction}
\author{Xinlong Dong}
\affiliation{College of Physics and Electronic Information $\&$ Key Laboratory of Magnetic Molecules and Magnetic Information Materials of Ministry of Education $\&$ Research Institute of Materials Science, Shanxi Normal University, Taiyuan 030031, China}
\author{Xuemin Shen}
\affiliation{College of Physics and Electronic Information $\&$ Key Laboratory of Magnetic Molecules and Magnetic Information Materials of Ministry of Education $\&$ Research Institute of Materials Science, Shanxi Normal University, Taiyuan 030031, China}
\author{Xiaowen Sun}
\affiliation{College of Physics and Electronic Information $\&$ Key Laboratory of Magnetic Molecules and Magnetic Information Materials of Ministry of Education $\&$ Research Institute of Materials Science, Shanxi Normal University, Taiyuan 030031, China}
\author{Yuhao Bai}
\affiliation{College of Physics and Electronic Information $\&$ Key Laboratory of Magnetic Molecules and Magnetic Information Materials of Ministry of Education $\&$ Research Institute of Materials Science, Shanxi Normal University, Taiyuan 030031, China}
\author{Zhi Yan}
\email[Corresponding author:~~]{yanzhi@sxnu.edu.cn}
\affiliation{College of Physics and Electronic Information $\&$ Key Laboratory of Magnetic Molecules and Magnetic Information Materials of Ministry of Education $\&$ Research Institute of Materials Science, Shanxi Normal University, Taiyuan 030031, China}
\affiliation{School of Chemistry and Materials Science of Shanxi Normal University $\&$ Collaborative Innovation Center for Shanxi Advanced Permanent Magnetic Materials and Technology, Taiyuan 030031, China}
\author{Xiaohong Xu}
\email[Corresponding author:~~]{xuxh@sxnu.edu.cn}
\affiliation{College of Physics and Electronic Information $\&$ Key Laboratory of Magnetic Molecules and Magnetic Information Materials of Ministry of Education $\&$ Research Institute of Materials Science, Shanxi Normal University, Taiyuan 030031, China}
\affiliation{School of Chemistry and Materials Science of Shanxi Normal University $\&$ Collaborative Innovation Center for Shanxi Advanced Permanent Magnetic Materials and Technology, Taiyuan 030031, China}

\date{\today{}}

\begin{abstract}
Multiferroic tunnel junctions (MFTJs) based on two-dimensional (2D) van der Waals heterostructures with sharp and clean interfaces at the atomic scale are crucial for applications in nanoscale multi-resistive logic memory devices. The recently discovered sliding ferroelectricity in 2D van der Waals materials has opened new avenues for ferroelectric-based devices. Here, we theoretically investigate the spin-dependent electronic transport properties of Fe$_3$GeTe$_2$/graphene/bilayer-$h$-BN/graphene/CrI$_3$ (FGT/Gr-BBN-Gr/CrI) all-vdW MFTJs by employing the nonequilibrium Green’s function combined with density functional theory. We demonstrate that such FGT/Gr-BBN-Gr/CrI MFTJs exhibit four non-volatile resistance states associated with different staking orders of sliding ferroelectric BBN and magnetization alignment of ferromagnetic free layer CrI$_3$, with a maximum tunnel magnetoresistance (electroresistance) ratio, i.e., TMR (TER) up to $\sim$$3.36\times10^{4}$\% ($\sim$$6.68\times10^{3}$\%) at a specific bias voltage. Furthermore, the perfect spin filtering and remarkable negative differential resistance effects are evident in our MFTJs. We further discover that the TMR, TER, and spin polarization ratio under an equilibrium state can be enhanced by the application of in-plane biaxial strain. This study highlights the significant potential of sliding ferroelectric BBN-based MFTJs in nonvolatile memories, showcasing their giant tunneling resistance ratio, multiple resistance states, and excellent spin-polarized transport properties.

\end{abstract}

\maketitle
\section{Introduction}
As the current increasing demand for faster, smaller, and non-volatile electronics, traditional silicon-based semiconductor devices are being pushed to smaller sizes. However, power dissipation and finite-size effects are limiting factors in device miniaturization. In this context, several new concepts for next-generation information processing and storage devices have been proposed and studied in recent years~\cite{xu2017correlated,kang2017layer,lin2019two,gong2019two}.
Of these, spintronics based on two-dimensional (2D) materials with only one or a few atomic layers offers broader possibilities for the miniaturization of emerging electronic devices with sophisticated functions. Especially, the 2D van der Waals (vdW) multiferroic tunneling junction (MFTJ) device stands out for its important multifunctional technological applications, which combine the tunneling magnetoresistance (TMR) effect of magnetic tunnel junctions (MTJs)~\cite{MTJ0, MTJ1, MTJ2, MTJ3, MTJ4, MTJ5, MTJ6, MTJ7, MTJ8, Li2021PRA} and the tunneling electroresistance (TER) effect of ferroelectric tunnel junctions (FTJs)~\cite{FTJ0, FTJ1, FTJ2, FTJ3}.

The coexistence of ferroelectric and magnetic order is crucial for MFTJs. However, the intrinsic single-phase multiferroics face a significant challenge due to the conflicting preferences of different ferroics with respect to the $d$-orbital occupation of metal ions. Ferroelectricity, which arises from the off-center positioning of cations, requires empty $d$-orbitals, whereas ferromagnetism usually results from partially filled $d$-orbitals~\cite{hill2000there}. In comparison to single-crystal multiferroics, 2D multiferroic vdW heterostructures are a more promising approach for designing MFTJs due to the weaker vdW interactions between the different layers. By employing this strategy, the first reported vdW MFTJs are constructed with a heterojunction composition of Fe$_n$GeTe$_2$/In$_2$Se$_3$($n$ = 3, 4, 5)~\cite{Su2020NL}. Later, we designed vdW MFTJs with six nonvolatile resistive states based on three ferroelectric polarization arrangements of bilayer $\upalpha$-In$_2$Se$_3$~\cite{yan2022giant}. Subsequently, Hu et al. proposed a Sc$_2$CO$_2$/CrI$_3$-based MFTJ and found a large TER and a low resistance-area (RA)~\cite{hu2023interface}. However, the ferroelectric polarization reversal in these MFTJs is reliant on an external electric field, resulting in unnecessary energy consumption.
Recently, a novel type of room-temperature ferroelectricity was both theoretically predicted~\cite{li2017binary,wu2021two} and experimentally demonstrated~\cite{FEBN} by sliding the parallel-stacking bilayer $h$-BN (BBN). This newly discovered ferroelectric mechanism offers a promising strategy for developing low-power ferroelectric storage. Yang et al. have reported sliding ferroelectric $h$-BN-based ferroelectric tunnel junctions~\cite{yang2022giant}, but as of now, MFTJs designed using this mechanism remain unexplored.
Another important factor for MFTJs is the selection of the magnetic layer, which directly impacts the device performance. Typically, a large TMR necessitates a highly spin-polarized magnetic layer, while a large TER requires asymmetric on both sides of the ferroelectric barrier layer. Therefore, here we choose two different magnetic materials, i.e., Fe$_3$GeTe$_2$ (FGT)\cite{FGT0Tc} and CrI$_3$~\cite{mcguire2015coupling}, as the magnetic layer. The smaller magnetocrystalline anisotropy of CrI$_3$~\cite{zhao2022enhanced} makes it suitable as the magnetic free layer, while the metalized FGT can serve as the ferromagnetic electrode.

In this work, we design MFTJs using the sliding ferroelectric bilayer $h$-BN (BBN) as the tunnel barrier and evaluate the corresponding device transport properties by using first-principles calculations.
We find a giant TMR ratio up to $\sim$$3.36\times10^{4}$\% at a bias voltage of -0.2 V by changing the magnetic orientation from a parallel (P) to antiparallel (AP) arrangement and a large TER ratio up to $\sim$$6.68\times10^{3}$\% at a bias voltage of 0.2 V by changing the stacking order of BBN (from AB to BA). Also, we observe a perfect spin filtering effect and a remarkable negative differential resistance effect. These results suggest that the four non-volatile resistance states can exist in sliding ferroelectric BBN-based MFTJs, which provides a promising platform for exploring novel atomic-scale spintronics.  

\begin{figure*}[htp!]
	\centering
	\includegraphics[width=17.5cm,angle=0]{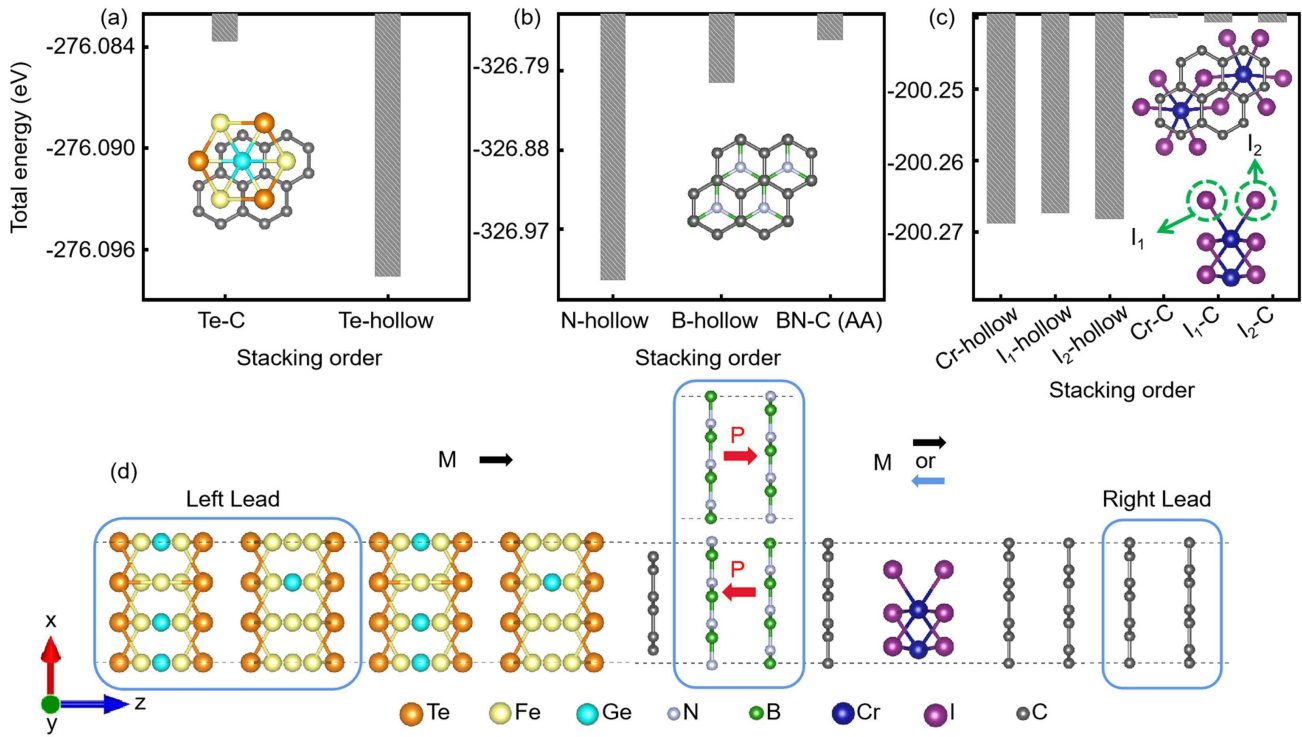}
	\caption{(a)-(c) The total energy of FGT/$h$-BN/CrI-Gr interfaces $vs$ the various stacking orders. The inset diagram represents the stack with the lowest energy for the corresponding interface circled in red. (d) Schematic diagrams of FGT/Gr-BBN-Gr/CrI MFTJs devices with AB- and BA-stacking BBN using metallic Fe$_3$GeTe$_2$ as the left lead and graphite as the right lead. The left and right lead extend to $\mp\infty$. These devices are periodic in the $xy$ plane and the current flows in the $z$ direction.}
	\label{Fig1}
\end{figure*}

\section{Computational Methods}
All structural relaxation calculations were performed within the framework of density functional theory as implemented in Vienna \textit{ab initio} simulation package (VASP)~\cite{VASP1996}. The electron-core interaction was described by the projected augmented wave~\cite{PAW1994}  pseudopotential with the general gradient approximation (GGA) parameterized by Perdew, Burke, and Ernzerhof (PBE)~\cite{GGA1992}. The electron wave function was expanded in plane waves up to a cutoff energy of 500 eV, and a $9\times9\times1$ Monkhorst–Pack $k$-grid~\cite{1976Special} was used to sample the Brillouin zone of the supercell. The convergence criterion for the electronic energy was set to be $10^{-5}$ eV. The Hellman–Feynman force was smaller than 0.01 eV/{\AA} in the optimized structure. A vacuum region of 20 {\AA} was used in the structural relaxation calculations of the central regions to avoid interaction between adjacent slabs. Moreover, the vdW force was considered through the DFT-D3 method in our calculations~\cite{DFTD3}. 

The calculations of quantum transport were calculated by using density functional theory coupled with the nonequilibrium Green's function (NEGF)~\cite{NEGF2001} as implemented in the Nanodcal package \cite{Nanodcal2001}.
The valence electrons are expanded in a numerical atomic-orbital basis set of double-zeta polarization (DZP) for all atoms\cite{double-zeta}. 
The cutoff energy was set to be 80 Hartree in our case, and the temperature
in the Fermi function was set to 300 K. In self-consistent calculation, the energy convergence criterion of the Hamiltonian matrix was set to be $10^{-5}$ eV. More $100\times100\times1$ $k$-mesh points were set for calculating the spin-reversed current and transmission coefficients of all MFTJs.

The spin-resolved current $I_\sigma$ and conductance $G_\sigma$ are calculated by using the Landauer-B\"uttiker formula \cite{CalcuI11992,book1995}:
\begin{align}
	I_\sigma &= \dfrac{e}{h} \int T_\sigma(E)[f_\text{L}(E) - f_\text{R}(E)] \mathrm{d}E,
\end{align}

\begin{align}
	G_\sigma &= \dfrac{e^{2}}{h}T_\sigma
\end{align}
where $\sigma$ represents the spin index ($\uparrow,\downarrow$), $e$ denotes the electron charge, $h$ is Planck's constant, $T_\sigma(E)$ is the spin-resolved transmission coefficient, and $f_\text{L(R)}(E)$ is the Fermi-Dirac distribution function of the left (right) electrode. The formula $I = I_\uparrow + I_\downarrow$  is utilized to calculate the total charge current $I$. The spin injection efficiency ($\eta$) is defined as:
\begin{align}
	\eta &=\left|\dfrac{I_\uparrow - I_\downarrow}{I_\uparrow + I_\downarrow}\right|.
\end{align}
In equilibrium state, the TMR ratio can be calculated by~\cite{TMR1C}:
\begin{align}
	\rm TMR &=\frac{G_\text{P}-G_\text{AP}}{G_\text{AP}}=\frac{T_\text{P}-T_\text{AP}}{T_\text{AP}},
\end{align}
at bias voltage $V$, 
\begin{align}
	 \text{TMR}_{(V)} &=\frac{I_\text{P}-I_\text{AP}}{I_\text{AP}},
\end{align}
where $T_\text{P/AP}$ and $I_\text{P/AP}$ are the total transmission coefficient of Fermi level and currents of bias voltage $V$ across the junctions in parallel ($\rm P$) and antiparallel ($\rm AP$) magnetic states, respectively.
Similarly, TER ratio is defined as \cite{TERCalcu2016,FTJ0,FTJ1}
\begin{align}
	\rm TER &=\frac{|G_\uparrow - G_\downarrow|}{\text{min}(G_\uparrow, G_\downarrow)}=\frac{|T_\uparrow - T_\downarrow|}{\text{min}(T_\uparrow, T_\downarrow)},
\end{align}
at bias voltage $V$, 
\begin{align}
	 \text{TER}_{(V)} &=\frac{|I_\uparrow - I_\downarrow|}{\text{min}(I_\uparrow, I_\downarrow)},
\end{align}
where $T_{\uparrow/\downarrow}$ and $I_{\uparrow/\downarrow}$ are the total transmission coefficient at the Fermi level and currents of bias voltage $V$ which can be obtained by reversing the direction of the ferroelectric polarization of the barrier layer.

\begin{figure}[htb!]
	\centering
	\includegraphics[width=8.5cm,angle=0]{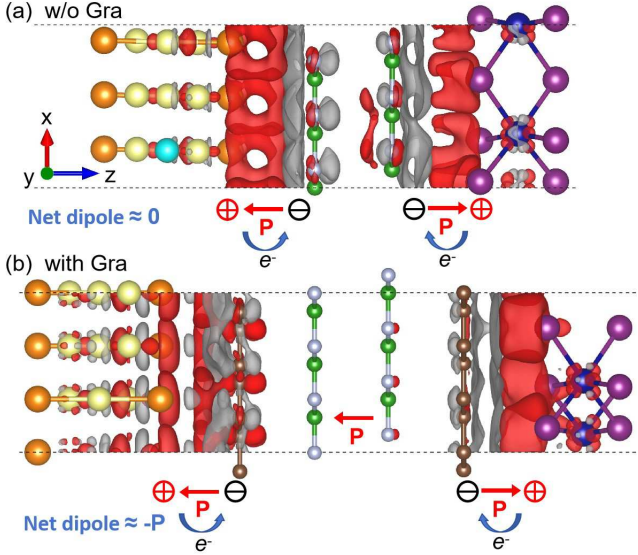}
	\caption{The difference charge density and interface dipole diagram for the central scattering region of the multiferroic tunnel junction. (a) without monolayer graphene; (b) with monolayer graphene. The isosurface value is set to be 0.0001 $e$/Bohr$^3$. Red and grey colors denote charge accumulation and depletion. The formation of the dipole is illustrated by the electron transfer. The symbols ‘+’ and ‘-’ represent the positive and negative charges, respectively.}
	\label{Fig2}
\end{figure}

\begin{figure}[htb!]
	\centering
	\includegraphics[width=8.5cm,angle=0]{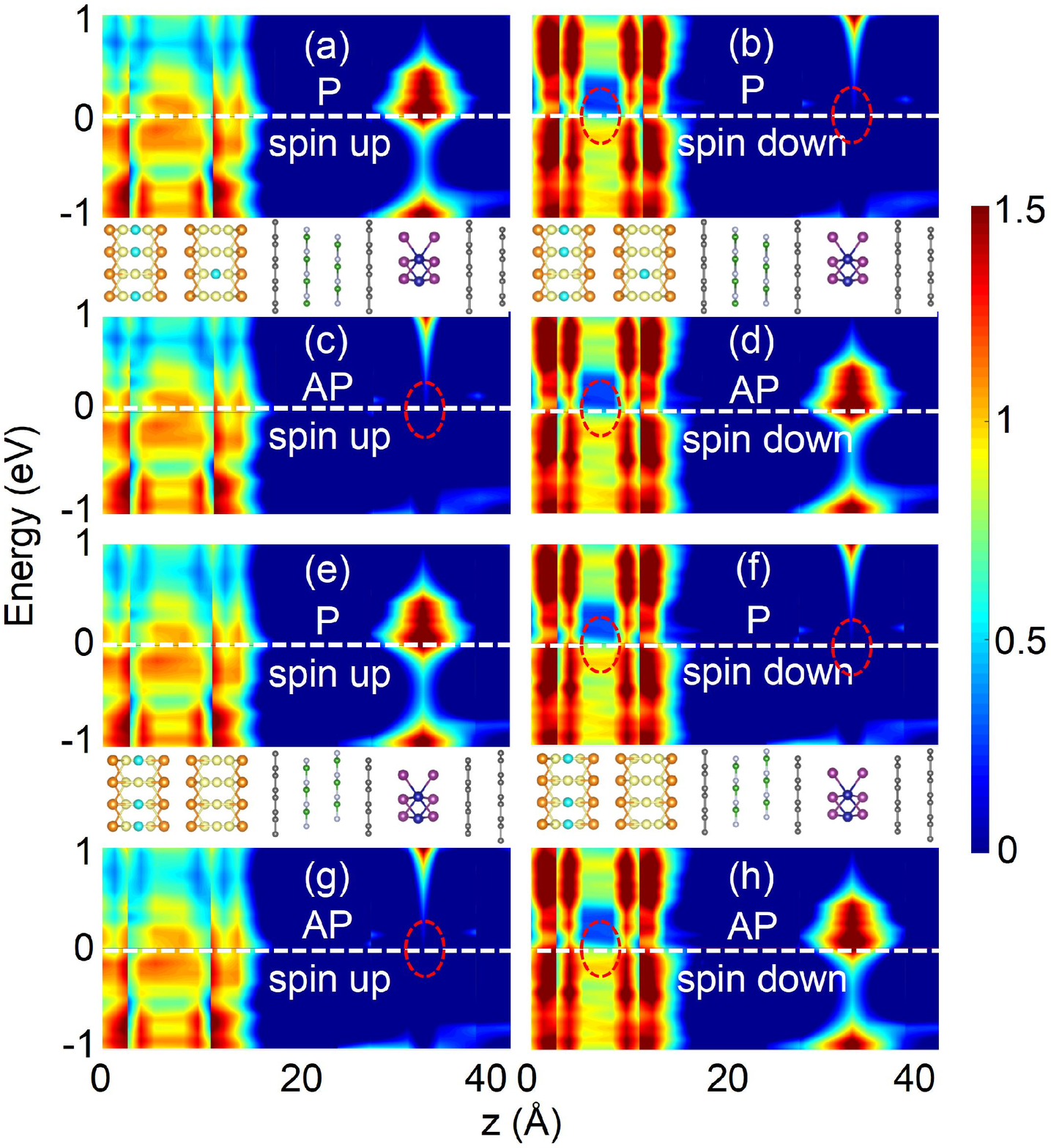}
	\caption{The averaged spin-resolved PDOS distribution over the $xy$ plane of the central scattering regions along the transport direction ($z-$axis) for FGT/Gr-BBN-Gr/CrI MFTJ in the equilibrium state. The white horizontal dashed line represents the Fermi level, and red dotted ellipses mark the minority electronic states. (a)-(d) AB-staking; (e)-(h) BA-stacking.}
	\label{Fig3}
\end{figure}

\section{Results and discussion}
\subsection*{A. The atomic model of MFTJs}
The AB(BA)-stacking bilayer $h$-BN (BBN) possesses a spontaneous out-of-plane ferroelectric polarization, which originates from the interlayer net charge transfer introduced by crystal asymmetry~\cite{li2017binary}. To confirm the ferroelectricity of BBN, we calculated the ferroelectric polarization to be $2.1\times10^{-12}$ Cm$^{-1}$ based on the Berry phase method~\cite{king1993theory}, which is consistent with previous study~\cite{li2017binary}. For AB(BA)-stacking BBN, the N(B) atoms in the upper layer are positioned directly above the B(N) atoms in the lower layer, while the B(N) atoms in the upper layer are situated above the vacant site at the center of the hexagon in the lower layer. This arrangement causes the 2$p_z$ orbitals of N and B to vertically align, resulting in distortion of the N orbital and generating an electric dipole moment. Therefore, the AB- and BA- stacking BBN exhibit opposite out-of-plane ferroelectric polarization directions.

\begin{table*}[htp!]
	\centering
	\renewcommand{\arraystretch}{1.8}
	\caption{Calculated spin-dependent electron transmission $T_{\uparrow}$ and $T_{\downarrow}$, TMR, TER, and spin injection efficiency $\eta$ at the equilibrium state for FGT/Gr-BBN-Gr/CrI MFTJs.}
	\label{table1}
	\begin{tabular}{c c c c c c c c c c c}
		\hline\hline
		\multicolumn{2}{c}{\raisebox{-0.8em}{Configuration}} & \multicolumn{4}{c}{P state (M${\uparrow\uparrow}$)} & \multicolumn{4}{c}{AP state (M${\uparrow\downarrow}$)} & \multirow{2}{*}{\large$\substack{\text{TMR}\\}$} \\ 
		\cline{3-10}
		\multicolumn{2}{c}{\raisebox{0.8em}{and Ratio}} & $T_{\uparrow}$ & $T_{\downarrow}$ & $T_\text{tot}=T_{\uparrow}+T_{\downarrow}$ & $\eta$ & $T_{\uparrow}$ & $T_{\downarrow}$ & $T_\text{tot}=T_{\uparrow}+T_{\downarrow}$ & $\eta$ &  \\
		\hline
		\multicolumn{2}{c}{AB-stacking} & $3.87\times10^{-6}$ & $3.74\times10^{-5}$ & $4.13\times10^{-5}$ & 81\% & $2.58\times10^{-5}$ & $8.45\times10^{-7}$ & $2.66\times10^{-5}$ & 94\% & 55\% \\
		\hline
		\multicolumn{2}{c}{BA-stacking} & $5.58\times10^{-6}$ & $3.87\times10^{-7}$ & $5.97\times10^{-6}$ & 87\% & $3.66\times10^{-7}$ & $1.67\times10^{-6}$ & $2.04\times10^{-6}$ & 64\% & 193\% \\
		\hline		
           \multicolumn{
           2}{c}{TER}& \multicolumn{4}{c}{592\%} & \multicolumn{4}{c}{1204\%} &  \\
		\hline\hline
	\end{tabular}
\end{table*}

After confirming the ferroelectricity of BBN, we can build Fe$_3$GeTe$_2$/graphene/bilayer-$h$-BN/graphene/CrI$_3$ (FGT/Gr-BBN-Gr/CrI) all-vdW MFTJs.
The optimized in-plane lattice constants of monolayer Fe$_3$GeTe$_2$, CrI$_3$, $h$-BN, and graphene are 3.998 {\AA}~\cite{FGTlatticecon}, 6.867 {\AA}~\cite{CrI3latticecon}, 2.511 {\AA}~\cite{BNlatticecon}, and 2.466 {\AA}~\cite{yan2020dramatically} respectively, so the corresponding minimum matching supercells to build MFTJs are $\sqrt{3}\times\sqrt{3}$, $1\times1$, $3\times3$ and $3\times3$. Considering the research focus on sliding ferroelectricity, we take the in-plane lattice constant of BBN as the overall lattice constant of MFTJs. In this case, the in-plane lattice matching results in stretching of FGT, CrI$_3$ and graphene by 8.04\%, 8.82\%, and 1.75\%, respectively. However, it is unlikely to observe such large values of compressive strain in experimental conditions, where the lattices of electrodes and barriers would be mismatched due to weak vdW-type bonding~\cite{MTJ2}. Nonetheless, these values are crucial for our computational simulations to maintain periodic boundary conditions. Furthermore, we also investigate the effect of in-plane biaxial strain on the transport properties of these MFTJs in the following Section B. There are three types of heterojunction interfaces in the FGT/Gr-BBN-Gr/CrI MFTJ, i.e., FGT-Gr, $h$-BN-Gr, and CrI-Gr. Owing to the presence of inversion symmetry, the FGT/Gr interface exhibits two distinct stacking orders (Te-C and Te-hollow), the $h$-BN/Gr interface displays three stacking orders (N-hollow, B-hollow, and BN-C), and the CrI/Gr interface manifests six stacking orders (Cr-C, Cr-hollow, I$_{1/2}$-hollow, and I$_{1/2}$-C). In these arrangements, X-C (hollow) represents one of the X atoms (X=Te, Cr, I, B, and N) located on top of the C atoms (the center of the hexagonal lattice) of monolayer graphene, specifically stating that BN-C refers to $h$-BN stacked with graphene in AA configuration. The total energy evaluation for the above three different interfaces with various stacking orders is summarized in Figs.~\ref{Fig1}(a)-(c). One can find that Te/N/Cr-hollow are the energy-favorite configurations of FGT/$h$-BN/CrI-Gr interfaces, respectively, and the corresponding visualized structures are presented in the inset of Fig~\ref{Fig1}. After establishing the stacking order of the aforementioned interfaces, a comprehensive transport device model of the FGT/Gr-BBN-Gr/CrI MFTJ can be constructed by integrating them and performing overall sufficient structural relaxation, where graphite and FGT are used as electrodes, as shown in Fig~\ref{Fig1}(d).
Note that monolayer graphene is added on both sides of the ferroelectric barrier layer BBN in the central scattering region to protect its interlayer ferroelectric polarization.
In Fig~\ref{Fig2}(a), it can be observed that in the absence of single-layer graphene on both sides of BBN, electrons are transferred from FGT/CrI$_3$ to BBN, creating interface dipoles. 
Each $h$-BN layer receives 0.031/0.025 $e$$^-$ from the adjacent FGT/CrI$_3$ layer based on the Bader charge analysis\cite{Bader}, resulting in a net electron dipole close to zero for the whole system. The stronger coupling between the FGT/CrI$_3$ layer and \textit{h}-BN leads to negligible mutual charge transfer within the BBN itself compared to the coupling within the BBN. In this case, the spontaneous polarization in the intrinsic AB- and BA-stacking BBN is absent. As depicted in Fig~\ref{Fig2}(b), when graphene is inserted, it effectively decouples \textit{h}-BN from FGT/CrI$_3$, enabling the preservation of interlayer ferroelectric polarization in BBN. Therefore, single-layer graphene was inserted on both sides of the BBN to construct our MFTJs.

\subsection*{B. TMR/TER effect and biaxial strain influence in equilibrium state}
According to the physical mechanism of MTJ and FTJ, flipping the magnetization direction of the magnetic free layer CrI$_3$ can form two opposite magnetic alignments (P/AP) with the ferromagnetic electrode FGT, while the sliding ferroelectric BBN exhibits two polarization directions, implying that four combined states can be induced in the MFTJ.
First, we study the TMR and TER effects of the FGT/Gr-BBN-Gr/CrI vdW MFTJs in equilibrium state. As presented in \autoref{table1}, the TMR ratio depends on the ferroelectric polarization of BBN, with TMR values of 55\% and 193\% for AB-stacking and BA-stacking, respectively.
In addition to the TMR effect, the TER effect is also an important factor in evaluating the performance of the FGT/Gr-BBN-Gr/CrI MFTJs.
As is evident from \autoref{table1}, the sliding of $h$-BN changes the ferroelectric polarization direction (AB$\rightarrow$BA) resulting in a TER ratio of 592\% (1204\%) in the P (AP) magnetic state.
Hence, the FGT/Gr-BBN-Gr/CrI all-vdW MFTJ exhibits favorable TMR and TER ratios, suggesting significant potential for utilization in nonvolatile memory devices.

\begin{figure}[htb!]
	\centering	
	\includegraphics[width=8.5cm,angle=0]{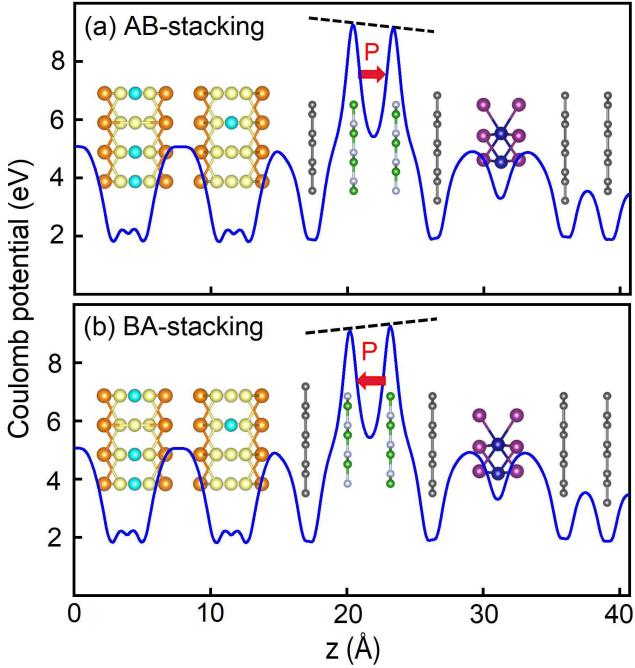}
	\caption{The evolution of the Coulomb potential in the central scattering region along the transport direction $z$ axis for the FGT/Gr-BBN-Gr/CrI vdW MFTJ. (a) AB-stacking; (b) BA-stacking.}
	\label{Fig4}
\end{figure}

\begin{figure}[htp!]
	\centering		
	\includegraphics[width=8.3cm,angle=0]{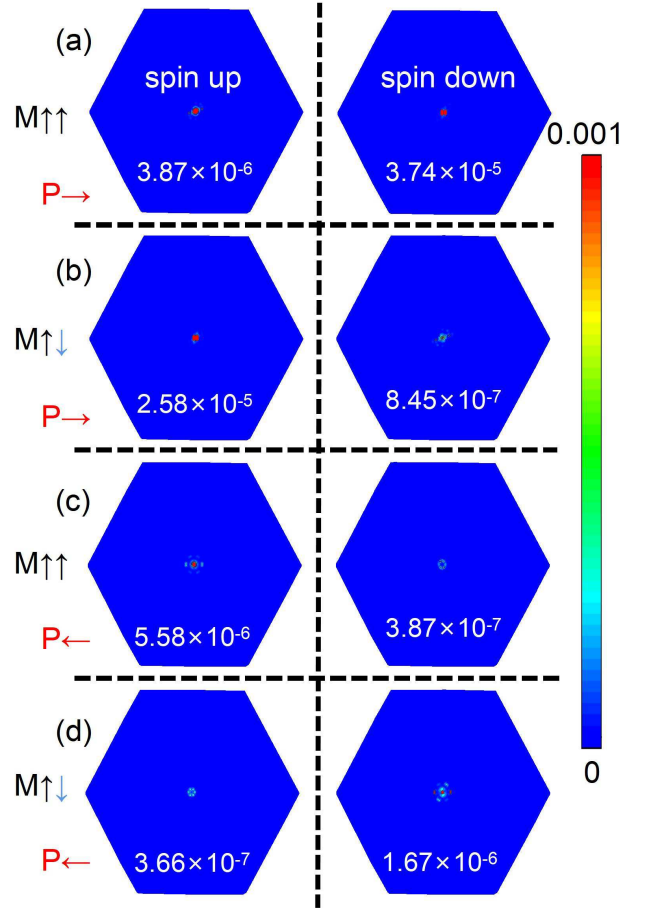}
	\caption{The $k_\Arrowvert$-resolved transmission coefficients across the FGT/Gr-BBN-Gr/CrI MFTJ in the 2D Brillouin zone at the Fermi level. (a)-(b) AB-stacking; (c)-(d) BA-stacking.}
	\label{Fig5}
\end{figure}

To clarify the various nonvolatile resistance states in FGT/Gr-BBN-Gr/CrI MFTJ, we conduct an analysis of the electronic structure of the central scattering region. In Fig.~\ref{Fig3}, we present the real-space projected density of states (PDOS) as a function of Fermi energy $E$ and vertical distance $z$ along the transport direction.
Obviously, the observed low electron states in the BBN region in the BBN region (specifically in the vicinity of positions from 21~{\AA} to 24~{\AA}) suggest that the BBN is sufficiently thick to serve as an insulating barrier and the tunneling mechanism governs that electron transport.
Furthermore, a typical TMR effect is clearly reflected in the PDOS diagram.
Since the MFTJs of the AB- and BA-stacking configurations exhibit similar PDOS  at $E=0$, here we choose the AB-stacking for detailed analysis.
Note the area marked with red dashed ellipses in Fig.~\ref{Fig3} is considered as a minority state because there is almost no electronic state distribution.
In Figs.~\ref{Fig3}(c)-(d), it can be observed that in the AP state, electrons with spin up (spin down) at the Fermi level move from the left FGT in the majority (minority) state, pass through the BBN ferroelectric barrier layer, and then traverse the right CrI$_3$ magnetic free layer in the minority (majority) state.
Undoubtedly, the opposite electronic state correspondence between FGT and CrI$_3$ on both sides of BBN hinders electron transport and may result in a high-resistance state.
In the P state, the spin up electrons are the majority states on both sides of BBN at the Fermi level, resulting in dominant transport with a low resistance state [refer to Fig.~\ref{Fig3}(a)].
Conversely, for the spin down state, there are two distinct dark blue regions, as indicated by red dashed ellipses in Fig.~\ref{Fig3}(b).
These regions exhibit an almost negligible DOS distribution near the Fermi level, hosting a high resistance state that may produce an outstanding spin-filtering effect, as summarized in \autoref{table1} (81\% for AB-stacking and 87\% for BA).
Figure.~\ref{Fig4} depicts the distribution of the Coulomb potential in the central scattering area of the MFTJ, showcasing the switchable built-in electric field within the BBN layer as indicated by the black dashed lines. This observation confirms the presence of ferroelectric polarization in the BBN interlayer.  Importantly, changing the stacking order of BBN effectively reverses the ferroelectric polarization direction, indicating the occurrence of the TER effect.

To provide a more comprehensive analysis of how ferroelectric polarization and magnetization alignment affect electron transmission, we perform calculations of the $k_\Arrowvert$-resolved transmission coefficients of the FGT/Gr-BBN-Gr/CrI MFTJs at the Fermi level within the 2D Brillouin zone (2D-BZ).
The results are depicted in Fig.~\ref{Fig5} for parallel (M$\uparrow\uparrow$) and antiparallel (M$\uparrow\downarrow$) magnetization alignments of the FGT and CrI$_3$ layers with the right (P$\rightarrow$) and left (P$\leftarrow$) ferroelectric polarizations of BBN. 
In 2D-BZ, minor transmissions can be observed near the $\Gamma$ point ($k_\Arrowvert$=0) for two stacked geometry MFTJs, indicating that the conductive channels only emerge at the $\Gamma$ point when electrons flow through single-layer graphene. This phenomenon can be attributed to the unique band folding through a special superlattice ($\sqrt{3}\times\sqrt{3}$) structure, resulting in the Dirac cone of graphene being located at the $\Gamma$ point.
Especially in the AP state, there are a significantly greater number of "hot spots" present in the spin up channel when compared to the spin down channel, with a difference in magnitude of over two orders of magnitude, as depicted in Fig.~\ref{Fig5}(b)]. This observation indicates a remarkable spin-filtering effect.
Moreover, comparing two stacked MFTJs with different ferroelectric polarization directions, denoted as P$\rightarrow$ and P$\leftarrow$, it is apparent that both P and AP states exhibit a significant order-of-magnitude difference in their electron transmission coefficients, resulting in a considerable TER ratio.
Conversely, these MFTJs with minor discrepancies in electron transmission coefficients between the P and AP states display a smaller TMR ratio.
Hence, the distribution of transmission coefficients in the 2D-BZ provides additional evidence to support the existence of a significant TER ratio and a considerable spin-filtering effect in these MFTJs in the equilibrium state.

\begin{figure}[htp!]
	\centering		
	\includegraphics[width=8.7cm,angle=0]{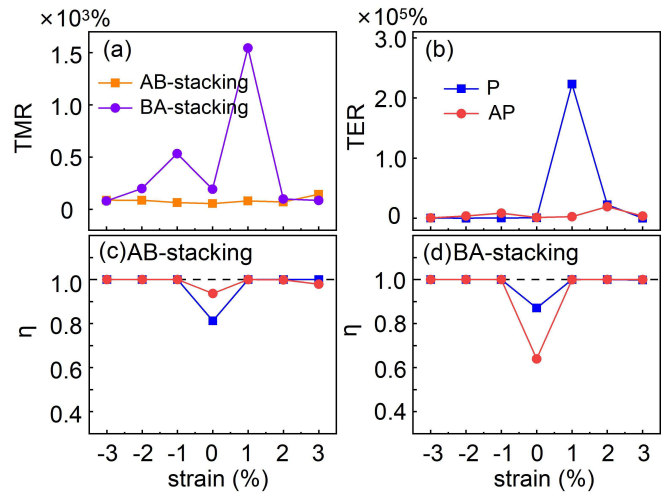}
	\caption{The (a) TMR, (b) TER, and (c), (d) spin injection efficiency $\eta$ of FGT/Gr-BBN-Gr/CrI MFTJs in different stacking orders as functions of in-plane biaxial strain.}
	\label{Fig6}
\end{figure}

\begin{figure*}[htp!]
	\centering	
	\includegraphics[width=18cm,angle=0]{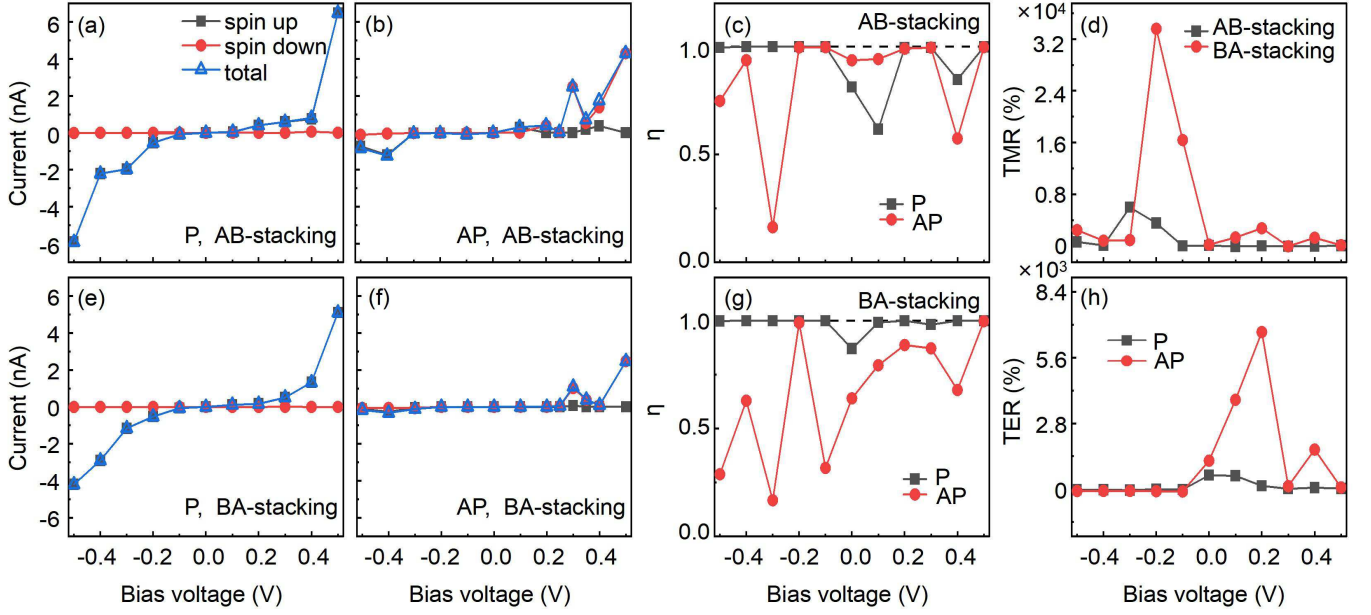}
	\caption{The variation of the current (a), (b), (e), and (f), spin injection efficiency $\eta$ (c) and (g), and TMR (d) and TER ratios (h) $vs$ the bias voltages for FGT/Gr-BBN-Gr/CrI MFTJs in different stacking orders. }
	\label{Fig7}
\end{figure*}

\begin{figure}[htp!]
	\centering	
	\includegraphics[width=8cm,angle=0]{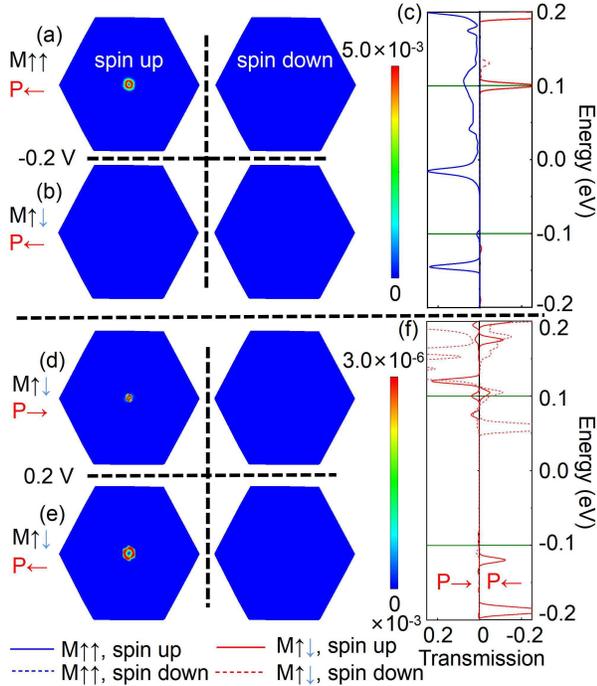}
	\caption{The $k_\Arrowvert$-resolved transmission coefficients across the FGT/Gr-BBN-Gr/CrI MFTJ in the 2D Brillouin zone and transmission coefficients as a function of energy at specific bias voltages that harbor the maximum TMR and TER. The dark green line represents the bias voltage window used to calculate the current. (a)-(c) -0.2 V; (d)-(f) 0.2 V.}
	\label{Fig8}
\end{figure}

Previous study demonstrated that stress is an effective means of modulating the transport properties of MFTJs~\cite{yan2022giant}. Considering the interface mismatch between different regions of these devices, we investigate the influence of in-plane biaxial strain on the transport behavior in equilibrium state as shown in Fig~\ref{Fig6}. 
We can observe that (i) the TMR ratio of BA-stacking MFTJ is adjustable and can be enhanced up to 1545\% under biaxial stress (1.0\% tensile strain) as displayed in Fig~\ref{Fig6}(a), while the TMR ratio of AB-stacking MFTJ remains stable within the investigated strain range; (ii) as shown in Fig~\ref{Fig6}(b), for the AP state, TER remains stable within the stress range, whereas for the P state, it remains stable under compressive stress and can be maximally increased up to $\sim$$2.23\times10^{5}$\% under a tensile stress of 1.0\%; (iii) as shown in Figs~\ref{Fig6}(c) and (d), stress can effectively enhance the spin injection efficiency for both AB- and BA-stacking MFTJs, leading to a perfect spin filtering effect at equilibrium state.
Hence, the above results suggest that strain is an effective approach for modulating the transport properties of MFTJs.

\subsection*{C. Voltage-tunable transport properties in nonequilibrium state}
Next, we calculate the current, spin polarization ratio ($\eta$) for the P and AP states, TMR, and TER ratio of the FGT/Gr-BBN-Gr/CrI MFTJ in the bias voltage range of -0.5 V to 0.5 V, as illustrated in Fig.~\ref{Fig7}.
Note that the bias voltage, $V{_b}$, is established by applying the chemical potential on the left (right) electrode as $+V_b/2$ ($-V_b/2$).
In the P magnetic order state, whether it is AB- or BA-stacking MFTJ, the total current shows a monotonically increasing trend with increasing bias voltage and is contributed by the spin-up current, implying a significant spin filtering effect [see Figs.~\ref{Fig7}(a) and (e)].
In the AP state of AB-stacking, as shown in Fig.~\ref{Fig7}(b), the total current first increases and then decreases in the bias ranges of 0.25 V to 0.35 V and -0.3 V to -0.5 V, exhibiting an interesting negative differential resistance effect.
Similarly, in the AP state of BA-stacking, a negative differential effect appears in the range of 0.25 V to 0.4 V [see Fig.~\ref{Fig7}(f)].
In addition, the spin injection efficiency can be tuned to almost 100\% by the bias voltage as seen in Figs.~\ref{Fig7}(c) and (g), indicating that the voltage can effectively manipulate the spin current.
Overall, the total current in the P state is much higher than that in the AP state, resulting in a larger TMR ratio at the corresponding bias voltage. Specifically, a TMR ratio as high as $\sim$$3.36\times10^{4}$\% is observed in the BA-stacking MFTJ when the bias voltage is -0.2 V as depicted in Fig.~\ref{Fig7}(d).
Another performance metric of the MFTJ device, TER, also depends on the bias voltage and can reach a maximum of $\sim$$6.68\times10^{3}$\% at 0.2 V for the AP state [see Fig.~\ref{Fig7}(h)].

To find out the large TMR, TER, and perfect spin filtering effect under the above special bias, we calculate the $k_\Arrowvert$-resolved transmission coefficients in the 2D-BZ and the electron transmission spectrums varying with energy at $\pm$0.2 V in Fig.~\ref{Fig8}.
Clearly, as shown in Fig.~\ref{Fig8}(a), in the P state (M$\uparrow\uparrow$), the "hot spots" only exist in the spin up channel around the $\Gamma$ point in the 2D-BZ, almost none in the spin down, which means that perfect spin filtering occurs at the Fermi level. While the AP state (M$\uparrow\downarrow$) hardly observed "hot spots", indicating a large TMR ratio as displayed in Fig.~\ref{Fig8}(b). More precisely, the current can be obtained by integrating the electron transmission spectra within the bias window according to Eq. (1). Therefore, the area enclosed by the electron transmission spectra within the energy axis of the bias window can approximately reflect the current. For the BA-stacking (P$\leftarrow$) MFTJ, only spin up transmission spectra exist within the bias window, regardless of the P or AP state, indicating perfect spin filtering effect, which is consistent with our calculated results of $\eta$ [see Fig.~\ref{Fig8}(c)]. Furthermore, the area within the bias window of the P state is significantly larger than that of the AP state from Fig.~\ref{Fig8}(c), implying a large TMR. As shown in Figs.~\ref{Fig8}(d) and (e), the "hot spots" of the AB-stacking (P$\rightarrow$) MFTJ in the 2D-BZ is much smaller than that of the BA-stacking (P$\leftarrow$) at the Fermi level, which may host a large TER, further confirmed by the area within the corresponding bias window [see Fig.~\ref{Fig8}(f)].

\section{Summary}
In conclusion, we theoretically study the spin-dependent electronic transport properties of the FGT/Gr-BBN-Gr/CrI all-vdW MFTJs based on the first-principles calculations. We demonstrate four non-volatile resistive states in MFTJ and the potential to manipulate these states through the application of bias or in-plane biaxial strain. The maximum achieved TMR (TER) is $3.36\times10^{4}$\% ($6.68\times10^{3}$\%) at a bias of -0.2V (0.2 V). Additionally, we observe perfect spin filtering and negative differential resistance effects. Our results reveal the potential of two-dimensional ferromagnets and sliding ferroelectrics for applications in ultrathin spin-electronics devices and future spintronics.

\section{Acknowledgements}
This work was supported by the National Key Research and Development Program of China (No. 2022YFB3505301), the Natural Science Basic Research Program of Shanxi (Nos. 20210302124252, 202203021222219, 202203021212393), and the Project funded by China Postdoctoral Science Foundation (No.2023M731452). Z. Y. thanks Xiao-wen Shi (from HZWTECH) and Wen-tian Lu for their valuable discussions.

\bibliography{main}
\bibliographystyle{apsrev4-2}
\end{document}